\documentclass[10pt,reprint,amsmath,amssymb,aps,pra,showpacs,longbibliography,superscriptaddress]{revtex4-1}
\usepackage[latin9]{inputenc}
\usepackage{mathrsfs}
\usepackage{amsmath}
\usepackage{amssymb}
\usepackage{graphicx}
\usepackage{esint}

\makeatletter

\usepackage{amsfonts}
\usepackage{graphics}

\makeatother

\begin{document}
\title{Thermally stable $p$-wave repulsive Fermi polaron without a two-body
bound state}
\author{Hui Hu}
\email{Corresponding author: hhu@swin.edu.au}

\affiliation{Centre for Quantum Technology Theory, Swinburne University of Technology,
Melbourne 3122, Australia}
\author{Jia Wang}
\affiliation{Centre for Quantum Technology Theory, Swinburne University of Technology,
Melbourne 3122, Australia}
\author{Xia-Ji Liu}
\affiliation{Centre for Quantum Technology Theory, Swinburne University of Technology,
Melbourne 3122, Australia}
\date{\today}
\begin{abstract}
We theoretically investigate the polaron physics of an impurity immersed
in a two-dimensional Fermi sea, interacting via a $p$-wave interaction
at finite temperature. In the unitary limit with a divergent scattering
area, we find a well-defined repulsive Fermi polaron at short interaction
range, which shows a remarkable thermal stability with increasing
temperature. The appearance of such a stable repulsive Fermi polaron
in the resonantly interacting limit can be attributed to the existence
of a quasi-bound dressed molecule state hidden in the two-particle
continuum, although there is no bound state in the two-particle limit.
We show that the repulsive Fermi polaron disappears when the interaction
range increases or when the scattering area is tuned to the weakly-interacting
regime. The large interaction range and small scattering area instead
stabilize attractive Fermi polarons.
\end{abstract}
\maketitle

\section{Introduction}

Fermi polaron is a well-established research area that addresses the
fate of an impurity when it moves and interacts with the surrounding
environment of a quantum many-fermion system \cite{Alexandrov2010}.
The ability of a free particle-like motion of the impurity under interactions
can be intuitively described by a fundamental concept of many-body
physics - quasiparticle \cite{Landau1933}. For many decades, the
determination of quasiparticle properties, such as energy, residue,
and lifetime, provides valuable insights on intriguing quantum many-fermion
systems in the different branches of physics \cite{Nozieres1969,Basile1990,Zhang2001,Chevy2006,Schirotzek2009,Sidler2017,Cao2022,Hu2023AB}.

Over the past two decades, the investigation of Fermi polarons with
ultracold atoms attracts particular interests \cite{Chevy2006,Schirotzek2009,Zhang2012,Kohstall2012,Koschorreck2012,Massignan2014,Scazza2017,Schmidt2018,Wang2023AB}.
Due to the unprecedented controllability on the interatomic interaction
(i.e., through Feshbach resonance \cite{Chin2010}), purity and dimensionality
of ultracold atomic Fermi gases \cite{Bloch2008,Hu2022AB}, new interesting
features of Fermi polarons can be revealed in a quantitative manner
\cite{Goulko2016,Wang2022PRL,Wang2022PRA}. For example, repulsive
Fermi polaron, which is an excited polaron state with well-defined
residue and lifetime, has been theoretically predicted \cite{Cui2010,Massignan2011}
and experimentally observed in an interacting imbalanced spin-1/2
Fermi gas with $s$-wave contact interactions \cite{Kohstall2012,Koschorreck2012,Scazza2017}.
This repulsive branch appears on the tightly bound side of the Feshbach
resonance, with a deep two-body bound state. Towards the strongly
interacting unitary limit at resonance where the two-body bound state
dissolves, the decay rate of repulsive Fermi polaron increases too
rapidly to behave like a well-defined quasiparticle \cite{Scazza2017,Massignan2011}.

In this work, we predict the existence of a well-defined repulsive
Fermi polaron in a two-dimensional Fermi gas, under a resonant $p$-wave
interaction between impurity and fermions, in the \emph{absence} of
a two-body bound state. This resonant unitary limit is described by
an infinitely large scattering area $a_{p}=\pm\infty$, together with
a nonzero effective interaction range $R_{p}$ \cite{Hu2018,Hu2019}.
The repulsive Fermi polaron appears at small interaction range, due
to a quasi-bound dressed molecule state hidden in the two-particle
continuum, although no two-body bound state can exist.

To describe the repulsive polaron branch, we apply a many-body $T$-matrix
theory \cite{Combescot2007,Hu2018FermiPolaron,Tajima2019,Hu2022}
that allows us to go beyond the earlier zero-temperature study of
$p$-wave Fermi polarons (in three dimensions) \cite{Levinsen2012}
and to explore their finite-temperature properties. Remarkably, the
$p$-wave repulsive Fermi polaron appears to be robust against thermal
fluctuations. This thermal stability of $p$-wave repulsive Fermi
polaron could be crucial for its experimental observation, since a
$p$-wave interacting Fermi gas often suffers severe atom loss at
low temperature \cite{Levinsen2008,Luciuk2016,Yoshida2018}. 

The rest of the paper is organized as follows. In the next section
(Sec. II), we outline the model Hamiltonian for the $p$-wave interacting
Fermi polaron and present the many-body $T$-matrix approach that
is capable to describe the one-particle-hole excitation at finite
temperature, which is the key ingredient of polaron physics. In Sec.
III, we first discuss the zero-momentum spectral function of the impurity
and show the appearance of the $p$-wave repulsive Fermi polaron in
the unitary limit. We then present the spectral function of molecule
to reveal the existence of a quasi-bound dressed molecule state. Finally,
we explore the parameter space for $p$-wave repulsive Fermi polaron,
by changing either the scattering area or the effective interaction
range. The conclusions follow in Sec. IV.

\section{Model Hamiltonian and many-body $T$-matrix approach}

We consider a highly imbalanced spin-1/2 Fermi gas of ultracold atoms
near a $p$-wave Feshbach resonance in two dimensions. In the extreme
limit of vanishing population of minority atoms, we treat minority
atoms as individual impurities, interacting with a non-interacting
Fermi sea of majority atoms via an interaction potential $V_{p}\left(\mathbf{k},\mathbf{k}'\right)$.
Our system can then be well-described by the single-channel model
Hamiltonian,

\begin{eqnarray}
\mathcal{H} & = & \sum_{\mathbf{k}}\xi_{\mathbf{k}}c_{\mathbf{k}}^{\dagger}c_{\mathbf{k}}+\sum_{\mathbf{k}}E_{\mathbf{k}}d_{\mathbf{k}}^{\dagger}d_{\mathbf{k}}\nonumber \\
 &  & +\sum_{\mathbf{kk'q}}V_{p}\left(\mathbf{k},\mathbf{k}'\right)c_{\frac{\mathbf{q}}{2}+\mathbf{k}}^{\dagger}d_{\frac{\mathbf{q}}{2}-\mathbf{k}}^{\dagger}d_{\frac{\mathbf{q}}{2}-\mathbf{k'}}c_{\frac{\mathbf{q}}{2}+\mathbf{k}'}^{\dagger},\label{eq:ModelHamiltonian}
\end{eqnarray}
where $c_{\mathbf{k}}^{\dagger}$ and $d_{\mathbf{k}}^{\dagger}$
are the creation field operators for majority atoms and the single
impurity, respectively. The first two terms in the Hamiltonian are
the single-particle terms with dispersion relations $\xi_{\mathbf{k}}=\hbar^{2}\mathbf{k}^{2}/(2m)-\mu=\epsilon_{\mathbf{k}}-\mu$
and $E_{\mathbf{k}}=\hbar^{2}\mathbf{k}^{2}/(2m)$, while the last
term describes the interaction term. Here, $\mu$ is the chemical
potential of majority atoms and at finite temperature $T$ is given
by, $\mu=k_{B}T\ln\left[\exp(\varepsilon_{F}/k_{B}T)-1\right]$. At
low temperature, the chemical potential approaches the Fermi energy
$\varepsilon_{F}=\hbar^{2}k_{F}^{2}/(2m)=2\pi n\hbar^{2}/m$, where
$n$ is the density of majority atoms in two dimensions.

For the $p$-wave interaction potential $V_{p}\left(\mathbf{k},\mathbf{k}'\right)$,
for concreteness we choose the chiral $p_{x}+ip_{y}$ channel, by
assuming that experimentally one can carefully tune the magnetic field
very close to a $p$-wave Feshbach resonance with azimuthal quantum
number $m=+1$. In practice, it is convenient to take a \emph{separable}
interaction potential in the form \cite{Hu2019,Botelho2005},
\begin{equation}
V_{p}\left(\mathbf{k},\mathbf{k}'\right)=\lambda\Gamma\left(\mathbf{k}\right)\Gamma\left(\mathbf{k}'\right),
\end{equation}
where $\lambda$ is the interaction strength and 

\begin{equation}
\Gamma\left({\bf k}\right)=\frac{\left(k/k_{F}\right)}{\left[1+\left(k/k_{0}\right)^{2n}\right]^{3/2}}e^{i\varphi_{{\bf k}}}
\end{equation}
is a dimensionless $p_{x}+ip_{y}$ form factor function with a cut-off
momentum $k_{0}$, a polar angle $\varphi_{{\bf k}}=\arctan(k_{y}/k_{x})$
and an exponent $n$ that is introduced for the convenience of numerical
calculations. The wavevector $k$ is measured in units of the Fermi
wavevector $k_{F}=(4\pi n)^{1/2}$.

\subsection{Two-body $T$-matrix and the renormalization of $p$-wave interaction}

While it is convenient to use the set of the three parameters ($\lambda,k_{0},n$)
to describe the $p$-wave interaction potential, physically we would
better use the scattering area $a_{p}$ (in two dimensions) and the
effective range of interaction $R_{p}$, which are formally defined
through the $p$-wave phase shift $\delta_{p}\left(k\right)$ in the
low-energy limit (i.e., $k\rightarrow0$) \cite{Hu2019,Levinsen2008},
\begin{equation}
k^{2}\cot\delta_{p}\left(k\right)=-\frac{1}{a_{p}}+\frac{2k^{2}}{\pi}\ln\left(R_{p}k\right)+\cdots.\label{eq:PhaseShift}
\end{equation}
The $p$-wave phase shift $\delta_{p}\left(k\right)$ can be easily
obtained by calculating the low-energy two-body $T$-matrix in vacuum
\cite{Hu2019}, $T_{2}^{(\textrm{vac})}({\bf k},{\bf k};E)=\left|\Gamma({\bf k})\right|^{2}\tilde{T}_{2}^{(\textrm{vac})}(E)$
with
\begin{equation}
\tilde{T}_{2}^{(\textrm{vac})}\left(E\right)=\left[\frac{1}{\lambda}+\sum_{{\bf p}}\frac{\left|\Gamma\left({\bf p}\right)\right|^{2}}{2\epsilon_{{\bf p}}-E-i0^{+}}\right]^{-1},\label{eq:T2_vacuum}
\end{equation}
where $E\equiv\hbar^{2}k^{2}/m$. Using the well-known relation $[T_{2}^{(\textrm{vac})}({\bf k},{\bf k};E)]^{-1}=-m[\cot\delta_{p}(k)-i]/(4\hbar^{2})$,
we find that in the low-energy limit $k\rightarrow0$,
\begin{equation}
\frac{1}{\lambda}+\mathcal{\mathscr{P}}\sum_{{\bf p}}\frac{\left|\Gamma\left({\bf p}\right)\right|^{2}}{2\epsilon_{{\bf p}}-E}=\frac{\left|\Gamma\left({\bf k}\right)\right|^{2}}{4E}\left[\frac{1}{a_{p}}-\frac{2k^{2}}{\pi}\ln\left(R_{p}k\right)\right],
\end{equation}
where $\mathscr{P}$ stands for taking Cauchy principal value. By
carrying out the summation over $\mathbf{p}$, we find that \cite{Hu2019}
\begin{eqnarray}
\frac{1}{a_{p}} & = & \frac{4\hbar^{2}k_{F}^{2}}{M\lambda}+\frac{\left(n-1/2\right)\left(n-1\right)}{n^{3}\sin\left(\pi/n\right)}k_{0}^{2},\\
R_{p} & = & \exp\left(\frac{3}{4n}\right)k_{0}^{-1}.
\end{eqnarray}
Throughout this work, we take an exponent $n=1.0$. For the given
scattering area $a_{p}$ and effective range $R_{p}$, we then determine
the interaction strength $\lambda$ and the cut-off momentum $k_{0}$
by using the above two expressions. Physical results of course will
not depend on the choice of the exponent $n$. In Appendix A, we explicitly
show the $n$-independence of the separable interaction potential
$V_{p}\left(\mathbf{k},\mathbf{k}'\right)$, by calculating the impurity
self-energy.

\subsection{Many-body $T$-matrix theory of Fermi polarons}

We now turn to solve the Fermi polaron problem by applying the many-body
$T$-matrix theory with the ladder approximation \cite{Combescot2007,Hu2018FermiPolaron,Tajima2019}.
The theory has been outlined in greater detail in the previous work
for Fermi polarons with an $s$-wave interaction potential \cite{Hu2022},
so here we only briefly sketch the key ingredients and emphasize the
changes due to the $p$-wave interaction potential. 

\begin{figure*}
\begin{centering}
\includegraphics[width=0.45\textwidth]{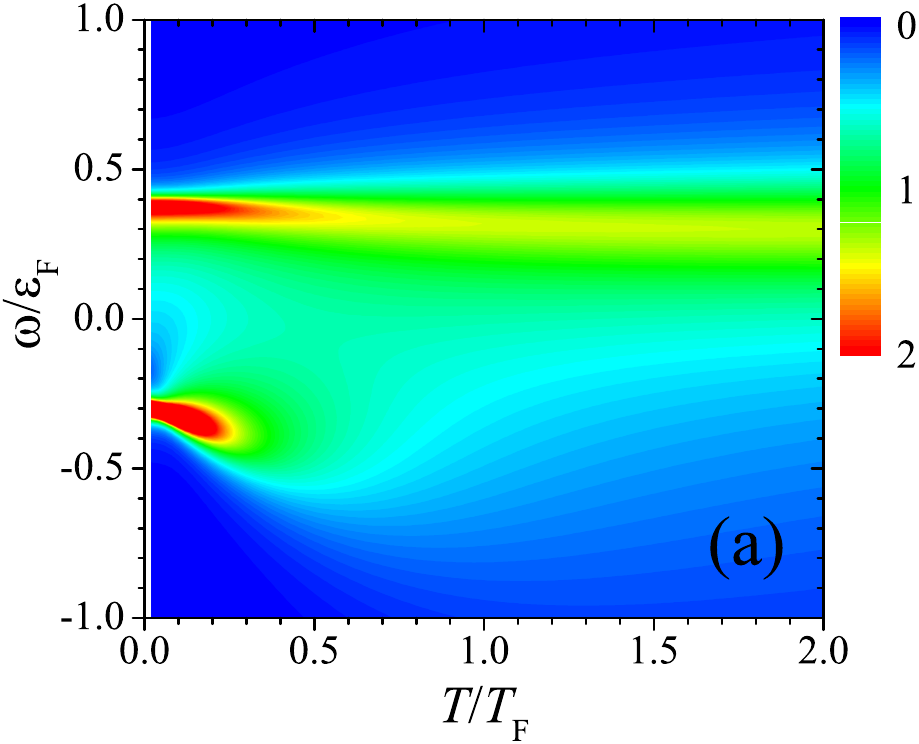}\includegraphics[width=0.475\textwidth]{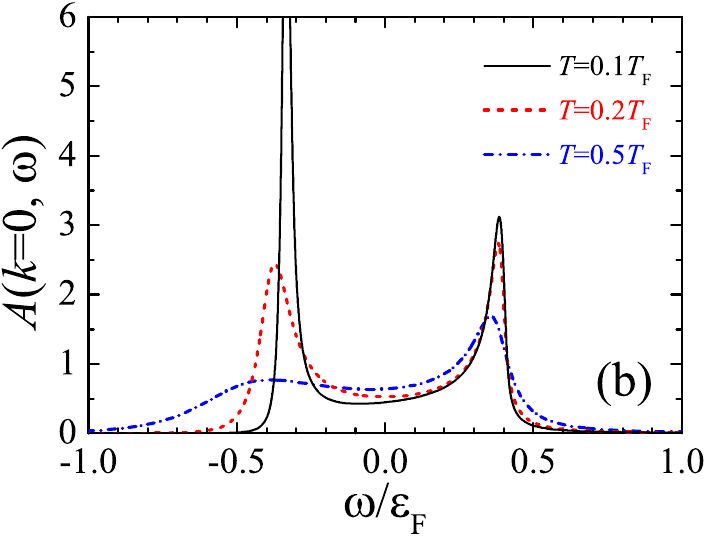}
\par\end{centering}
\centering{}\caption{\label{fig:fig1_akwkFRp010} Zero-momentum spectral function $A(\mathbf{k}=0,\omega)$
of the impurity as a function of the reduced temperature $T/T_{F}$.
The two-dimensional contour plot in (a) is shown in the linear scale.
The three lines in (b) show the spectral functions at $T=0.1T_{F}$
(black solid), $T=0.2T_{F}$ (red dashed) and $T=0.5T_{F}$ (blue
dot-dashed). Here, we consider the unitary limit $1/(k_{F}^{2}a_{p})=0$
and take an interaction range $k_{F}R_{p}=0.10$. The spectral functions
are measured in units of $\varepsilon_{F}^{-1}$.}
\end{figure*}

Following the same derivation as in the previous work \cite{Hu2022},
it is easy to show that the many-body $T$-matrix is given by
\begin{equation}
\tilde{T}_{2}({\bf k},{\bf k}',\mathbf{q};\omega)=\Gamma\left({\bf k}\right)\Gamma^{*}\left({\bf k'}\right)\tilde{T}_{2}\left(\mathbf{q},\omega\right),
\end{equation}
where
\begin{equation}
\tilde{T}_{2}^{-1}\left(\mathbf{q},\omega\right)=\frac{1}{\lambda}-\sum_{{\bf p}}\left|\Gamma\left({\bf p}\right)\right|^{2}\frac{1-f\left(\xi_{\frac{\mathbf{q}}{2}+\mathbf{p}}\right)}{\omega-\xi_{\frac{\mathbf{q}}{2}+\mathbf{p}}-E_{\frac{\mathbf{q}}{2}-\mathbf{p}}+i0^{+}}.\label{eq:T2_manybody}
\end{equation}
By taking $\mathbf{q}=0$, $\mu=0$ and the Fermi-Dirac distribution
function $f(x)=0$ (i.e., by considering the two-body scattering in
vacuum), $\tilde{T}_{2}(\mathbf{q};\omega)$ reduces to the two-body
$T$-matrix Eq. (\ref{eq:T2_vacuum}). In other words, the many-body
$T$-matrix can be understood as the effective interaction in the
medium. According to the interaction Hamiltonian in Eq. (\ref{eq:ModelHamiltonian}),
diagrammatically this effective interaction comes with two incoming
leges with momenta $\mathbf{q}/2\pm\mathbf{k}'$ and with two out-going
legs with momenta $\mathbf{q}/2\pm\mathbf{k}$. Therefore, by winding
back the out-going leg of majority atoms (i.e., with the momentum
$\mathbf{q}/2+\mathbf{k}$) and connecting it to the incoming leg
with the momentum $\mathbf{q}/2+\mathbf{k}'$ (so $\mathbf{k}=\mathbf{k}'$)
\cite{Hu2022}, we obtain the approximate impurity self-energy in
the many-body $T$-matrix theory,
\begin{equation}
\Sigma\left(\frac{\mathcal{Q}}{2}-\mathcal{K}'\right)=\sum_{\mathcal{Q}}\mathscr{G}\left(\frac{\mathcal{Q}}{2}+\mathcal{K}'\right)\tilde{T}_{2}\left(\mathcal{Q}\right)\left|\Gamma\left({\bf k}'\right)\right|^{2},
\end{equation}
where $\mathscr{G}$ is non-interacting Green function of the Fermi
sea and we use $\mathcal{Q}=(\mathbf{q},i\nu_{n})$ and $\mathcal{K}'=(\mathbf{k}',i\omega_{m'})$
to denote the four-dimensional momenta with the bosonic Matsubara
frequency $\nu_{n}=2\pi nk_{B}T$ and the fermionic Matsubara frequency
$\omega_{m'}=\pi(2m'+1)k_{B}T$ at finite temperature $T$. We have
also used the abbreviation $\sum_{\mathcal{Q}}\equiv\sum_{\mathbf{q}}k_{B}T\sum_{i\nu_{n}}$.
For the impurity self-energy, it is convenient to change to $\mathcal{K}=\mathcal{Q}/2-\mathcal{K}'=(\mathbf{k},i\omega_{m}$),
so $\Sigma(\mathcal{K})=\sum_{\mathcal{Q}}G(\mathcal{Q}-\mathcal{K})\tilde{T}_{2}(\mathcal{Q})\left|\Gamma(\mathbf{q}/2-{\bf k})\right|^{2}.$
By summing over the bosonic Matsubara frequency $\nu_{n}$ and taking
the analytic continuation $i\omega_{m}\rightarrow\omega+i0^{+}$ \cite{Hu2022},
we find the impurity self-energy,
\begin{equation}
\Sigma\left(\mathbf{k},\omega\right)=\sum_{\mathbf{q}}f\left(\xi_{\mathbf{q}-\mathbf{k}}\right)\tilde{T}_{2}\left(\mathbf{q},\omega+\xi_{\mathbf{q}-\mathbf{k}}\right)\left|\Gamma\left(\frac{\mathbf{q}}{2}-{\bf k}\right)\right|^{2}.\label{eq:SelfEnergy}
\end{equation}

In comparison with the case of an $s$-wave interaction potential
\cite{Hu2022}, the many-body $T$-matrix Eq. (\ref{eq:T2_manybody})
and the impurity self-energy Eq. (\ref{eq:SelfEnergy}) take essentially
the same forms, apart from the additional interaction form factors
$\Gamma({\bf k})$ and $\Gamma^{*}({\bf k'})$ that are necessary
to characterize the $p$-wave interaction \cite{Hu2019,Botelho2005}.
These interaction form factors do not introduce too many numerical
workloads, and the numerical difficulty still lies on the handling
of the pole structure that might appear in the summation over the
momentum $\mathbf{p}$ in $\tilde{T}_{2}^{-1}(\mathbf{q},\omega)$.
A detailed discussion of $\tilde{T}_{2}^{-1}(\mathbf{q},\omega)$
is included in Appendix B. We note that, the summation over the momentum
$\mathbf{q}$ in the self-energy $\Sigma(\mathbf{k},\omega)$ may
also suffer from the existence of a well-defined molecule state, which
manifests itself as a pole or a delta-peak of $\tilde{T}_{2}(\mathbf{q},\omega)$.

\section{Results and discussions}

Once the impurity self-energy is obtained, we directly calculate the
impurity Green function \cite{Massignan2014,Hu2022},
\begin{equation}
G\left(\mathbf{k},\omega\right)=\frac{1}{\omega-E_{\mathbf{k}}-\Sigma\left(\mathbf{k},\omega\right)}.
\end{equation}
Fermi polarons can be well-characterized by the impurity spectral
function 
\begin{equation}
A\left(\mathbf{k},\omega\right)=-\frac{1}{\pi}\textrm{Im}G\left(\mathbf{k},\omega\right),
\end{equation}
where each quasiparticle is visualized by the appearance of a sharp
spectral peak, with its width reflecting the lifetime or decay rate
of quasiparticle \cite{Massignan2011,Combescot2007}. Mathematically,
we may determine the quasiparticle energy $\mathcal{E}_{P}(\mathbf{k})$,
either the attractive polaron energy or the repulsive polaron energy,
from the pole position of the impurity Green function ($\omega\rightarrow\mathcal{E}_{P}(\mathbf{k)}$),
i.e., 
\begin{equation}
\mathcal{E}_{P}\left(\mathbf{k}\right)=E_{\mathbf{k}}+\textrm{Re}\Sigma\left[\mathbf{k},\mathcal{E}_{P}\left(\mathbf{k}\right)\right].\label{eq:PolaronEnergy}
\end{equation}
By expanding the self-energy near the zero momentum $\mathbf{k}=0$
and the polaron energy $\mathcal{E}_{P}\equiv\mathcal{E}_{P}\left(\mathbf{0}\right)$,
we directly determines various quasiparticle properties, including
the polaron residue,
\begin{equation}
\mathcal{Z}^{-1}=1-\left.\frac{\partial\textrm{Re}\Sigma(\mathbf{0},\omega)}{\partial\omega}\right|_{\omega=\mathcal{E}_{P}},
\end{equation}
and the polaron decay rate,
\begin{equation}
\Gamma=-2\mathcal{Z}\textrm{Im}\Sigma\left(\mathbf{0},\mathcal{E}_{P}\right).
\end{equation}

\begin{figure}
\begin{centering}
\includegraphics[width=0.5\textwidth]{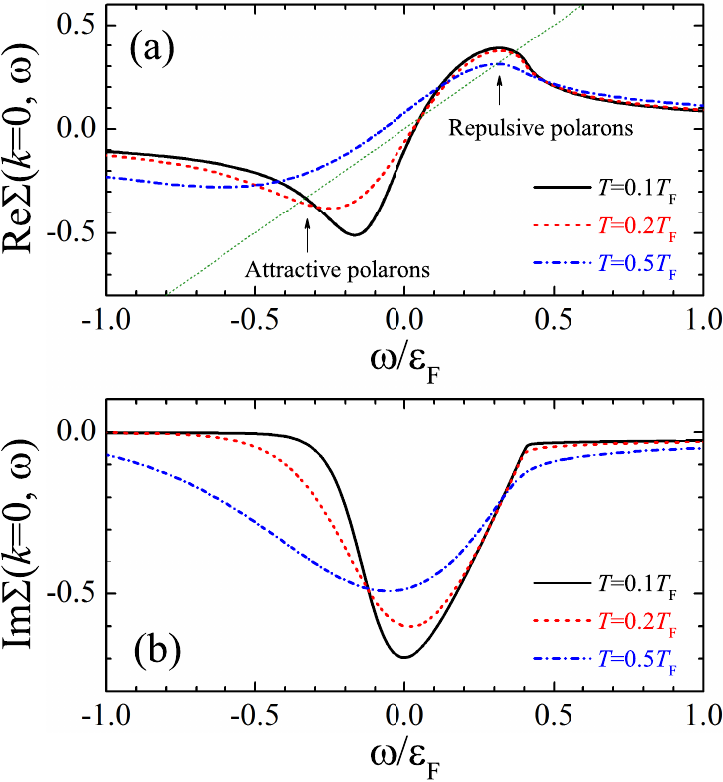}
\par\end{centering}
\centering{}\caption{\label{fig:fig2_SelfEnergykFRp010} Real part (a) and Imaginary part
(b) of the impurity self-energy as a function of the frequency $\omega$.
Both self-energy and frequency are measured in units of the Fermi
energy $\varepsilon_{F}$. The three lines show the self-energy at
$T=0.1T_{F}$ (black solid), $T=0.2T_{F}$ (red dashed) and $T=0.5T_{F}$
(blue dot-dashed). In (a), we also show the curve $y=\omega$ using
a dotted line. The crossing points of this curve with the three lines
for the real part of self-energy determine the polaron energies, which
are indicated using arrows for both attractive polaron branch and
repulsive polaron branch. As in Fig. \ref{fig:fig1_akwkFRp010}, we
consider the unitary limit $1/(k_{F}^{2}a_{p})=0$ and take an interaction
range $k_{F}R_{p}=0.10$. }
\end{figure}

\subsection{Repulsive Fermi polaron in the unitary limit}

Let us first focus on the unitary limit, where the scattering area
diverges, i.e., $a_{p}=\pm\infty$. In Fig. \ref{fig:fig1_akwkFRp010}(a),
we report the temperature evolution of the zero-momentum spectral
function $A(\mathbf{k}=0,\omega$) at a small interaction range $k_{F}R_{p}=0.1$,
in the form of a two-dimensional contour plot in the linear scale.
The one-dimensional plots of the spectral functions at the three typical
temperatures $T=0.1T_{F}$, $0.2T_{F}$ and $0.5T_{F}$ are shown
in Fig. \ref{fig:fig1_akwkFRp010}(b), by using black solid, red dashed,
and blue dot-dashed lines, respectively.

At very low temperature (i.e., $T\sim0$), we find two dominant peaks
in the spectral function at the positions $\omega\simeq-0.31\varepsilon_{F}$
and $\omega\simeq+0.37\varepsilon_{F}$, which correspond to the attractive
Fermi polaron and the repulsive Fermi polaron \cite{Goulko2016,Wang2022PRL,Wang2022PRA,Massignan2011},
respectively. By increasing temperature, both polaron states show
a red-shift in energy, similar to the $s$-wave case \cite{Hu2022}.
The initially sharp attractive polaron peak quickly dissolves into
a broad distribution at $T\sim0.3T_{F}$, and eventually disappears
at $T\sim0.6T_{F}$. In sharp contrast, the repulsive polaron appears
to be very robust against thermal fluctuations. In particular, once
the temperature is larger than $0.6T_{F}$, the position and the width
of the repulsive polaron peak essentially do not change with temperature.

To confirm the existence of a repulsive Fermi polaron, in Fig. \ref{fig:fig2_SelfEnergykFRp010}
we present the real part (a) and the imaginary part (b) of the self-energy
at the same three typical temperatures as in Fig. \ref{fig:fig1_akwkFRp010}(b).
At zero momentum (i.e., $E_{\mathbf{k}}=0$), a pole of the impurity
Green function occurs when $\omega=\textrm{Re}\Sigma(k=\mathbf{0},\omega)$.
Therefore, in Fig. \ref{fig:fig2_SelfEnergykFRp010}(a) the intercept
between the green dotted line (i.e., $y=\omega$) and the curves $\textrm{Re}\Sigma(k=\mathbf{0},\omega)$
determines the polaron energy. On the negative frequency side ($\omega<0$),
we find that the green dotted line fails to intercept with the curve
$\textrm{Re}\Sigma(k=\mathbf{0},\omega)$ at the temperate $T=0.5_{F}$,
consistent with our earlier observation that the attractive polaron
develops into a broad structure once $T>0.3T_{F}$. On the positive
frequency side ($\omega>0$), the green dotted line always crosses
with the curves $\textrm{Re}\Sigma(k=\mathbf{0},\omega)$ at different
temperatures, indicating the persistence of the repulsive polarons
with increasing temperature. The position of the crossing points or
the repulsive polaron energies do not change too much as temperature
increases. Remarkably, at those repulsive polaron energies, the imaginary
part of the self-energy $\textrm{Im}\Sigma(k=\mathbf{0},\omega)$
turns out to be reasonably small, and more importantly to be temperature
insensitive. 

The thermally stable repulsive polaron is not expected to appear in
the unitary limit, because a well-defined two-body bound state does
not exist \cite{Massignan2014,Massignan2011}. This situation might
be compared with an $s$-wave Fermi polaron in three dimensions \cite{Hu2022}.
In that case, in the same unitary limit, where the $s$-wave scattering
length $a_{s}$ diverges, the impurity spectral function only shows
an extremely broad background at positive energy, without any signal
for a repulsive polaron. The analysis of the real part of the self-energy
confirms the absence of a repulsive polaron, since no solution exists
for the condition $\omega=\textrm{Re}\Sigma(k=\mathbf{0},\omega)$
when $\omega>0$ \cite{Hu2022}. Under the $s$-wave interaction between
the impurity and the Fermi sea, repulsive polaron only appears in
the tightly bound limit, where a well-defined two-body bound state
exists. Towards the $s$-wave unitary limit, the decay rate of the
$s$-wave repulsive polaron will quickly increase with both the scattering
length $a_{s}$ and temperature. As a result, there is no meaningful
$s$-wave repulsive polaron near the strongly interacting regime of
the unitary limit.

\begin{figure*}
\begin{centering}
\includegraphics[width=0.33\textwidth]{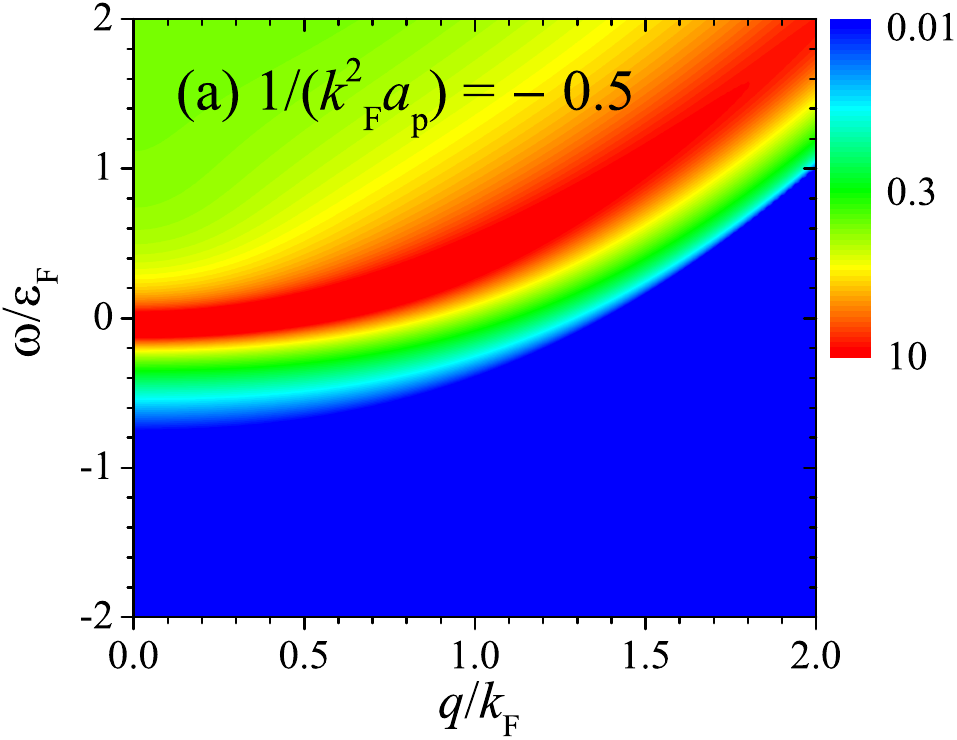}\includegraphics[width=0.33\textwidth]{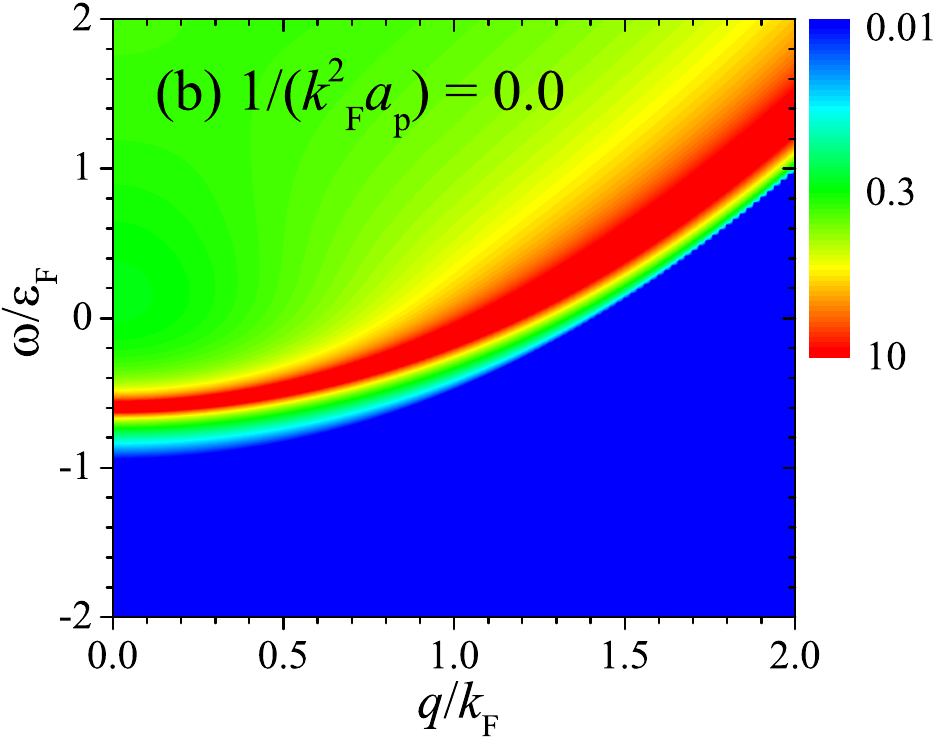}\includegraphics[width=0.33\textwidth]{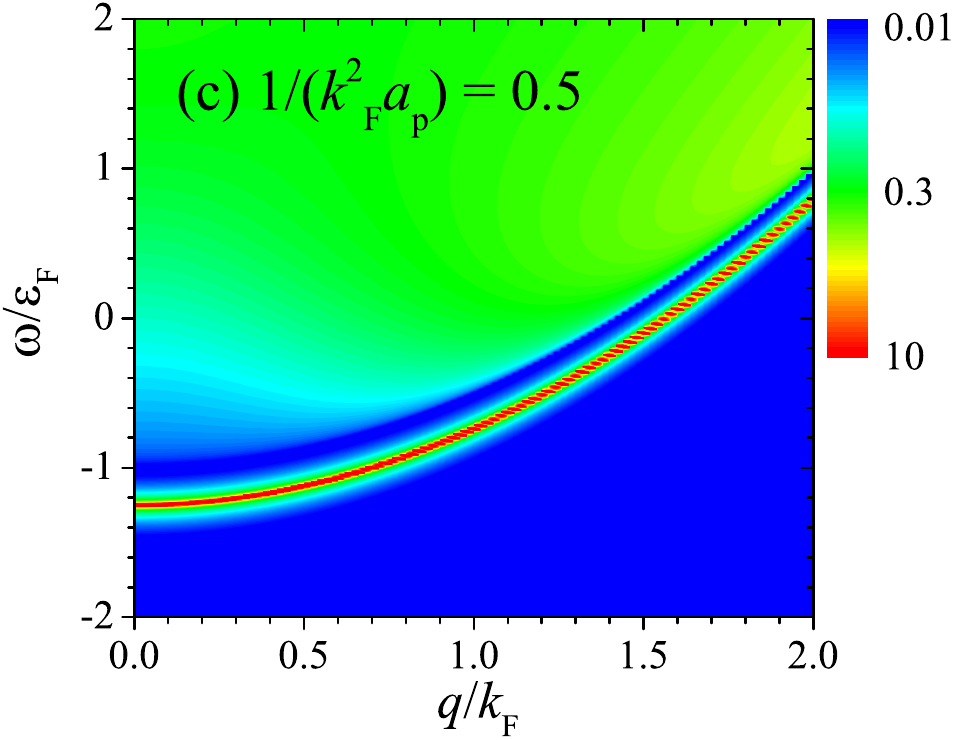}
\par\end{centering}
\centering{}\caption{\label{fig:fig3_moleculeAkw} Two-dimensional contour plots of the
molecule spectral function $A_{\textrm{mol}}(\mathbf{q},\omega)$
at different inverse scattering areas: $1/(k_{F}^{2}a_{p})=-0.5$
(a), $1/(k_{F}^{2}a_{p})=0$ (b) and $1/(k_{F}^{2}a_{p})=+0.5$ (c).
The contour plots are shown in the logarithmic scale in arbitrary
units, as indicated by the color bar. We take an interaction range
$k_{F}R_{p}=0.1$ at temperature $T=0.2T_{F}$.}
\end{figure*}

The existence of a thermally robust repulsive polaron, without a two-body
bound state, is therefore an unique feature of $p$-wave Fermi polarons.
To understand its formation mechanics, we consider the dressed molecule
state in the presence of the many-body environment of a Fermi sea,
which is an analogue of a Cooper pair in the limit of an extreme population
imbalance. In the many-body $T$-matrix theory, the dressed molecule
state is simply described by its effective Green function, i.e, the
many-body $T$-matrix $T_{2}(\mathbf{q},\omega)$ \cite{Combescot2007,Hu2018FermiPolaron,Hu2022}.
Thus, we introduce a molecule spectral function,
\begin{equation}
A_{\textrm{mol}}\left(\mathbf{q},\omega\right)=-\frac{1}{\pi}\textrm{Im}T_{2}\left(\mathbf{q},\omega\right).
\end{equation}
In Fig. \ref{fig:fig3_moleculeAkw}, we report the two-dimensional
contour plots of the molecule spectral function in the logarithmic
scale across the unitary limit at the temperature $0.2T_{F}$, with
the inverse scattering areas $1/(k_{F}^{2}a_{p})=-0.5$ (a), $0.0$
(b), and $+0.5$ (c). To be consistent with the results in Fig. \ref{fig:fig1_akwkFRp010},
we have taken the same effective interaction range $k_{F}R_{p}=0.1$. 

On the molecule side of the Feshbach resonance in Fig. \ref{fig:fig3_moleculeAkw}(c),
it is readily to see a sharp peak starting at energy $\omega\sim-1.2\varepsilon_{F}$,
which is well-separated from the two-particle continuum. Moreover,
at the momentum $q<k_{F}$ the spectral weight at the bottom of the
two-particle continuum is depleted. The separate, sharp peak arises
from of the existence of an \emph{undamped} dressed molecule state,
which in the two-particle limit reduces to a two-body bound state
that must exist with a positive scattering area. 

In the unitary limit on resonance (see Fig. \ref{fig:fig3_moleculeAkw}(b)),
although a separate undamped peak does not exist, we do observe that
a sharp peak emerges at energy $\omega\sim-0.6\varepsilon_{F}$ and
becomes broader when the momentum $q$ is larger than $0.5k_{F}$.
We would like to attribute the existence of a well-defined repulsive
polaron in the unitary limit to this quasi-bound dressed molecule
state, which is hidden slightly above the bottom of the two-particle
scattering continuum. Although it is a quasi-bound molecule state,
it effectively depletes fermions surrounding the dressed molecule
and eventually leads to the excited repulsive polaron \cite{Massignan2011}.
This idea is also supported by the weak temperature-dependence of
the quasi-bound dressed molecule state (not shown in the figure),
which may explain the observed thermal stability of the repulsive
polaron.

We finally consider a negative scattering area, as shown in Fig. \ref{fig:fig3_moleculeAkw}(a).
The dressed molecule state now becomes less well-defined, with a much
broader peak and with its energy blue-shifted to $\omega\sim0$ at
zero momentum. In this situation, the repulsive polaron may fail to
exist, as we shall see.

\begin{figure}
\begin{centering}
\includegraphics[width=0.5\textwidth]{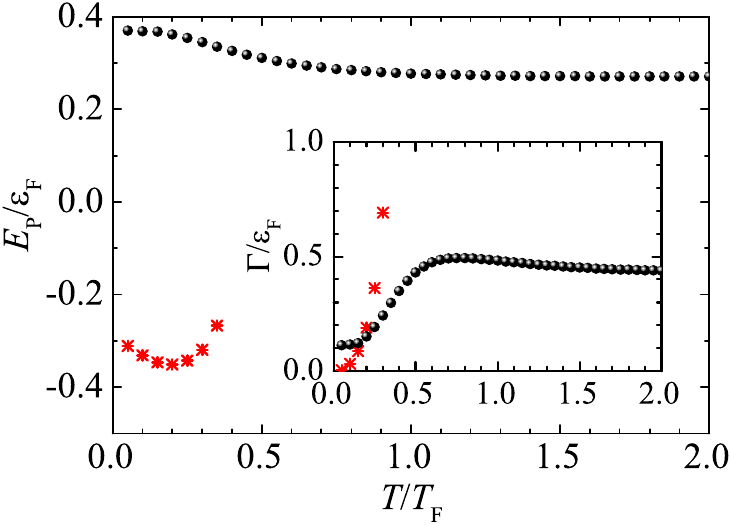}
\par\end{centering}
\centering{}\caption{\label{fig:fig4_PolaronkFRp010} Temperature dependence of the attractive
polaron energy (red stars) and of the repulsive polaron energy (black
dots) in the unitary limit $1/(k_{F}^{2}a_{p})=0$ with an interaction
range $k_{F}R_{p}=0.1$. The inset shows the temperature dependence
of the corresponding decay rates.}
\end{figure}

\begin{figure*}
\begin{centering}
\includegraphics[width=0.45\textwidth]{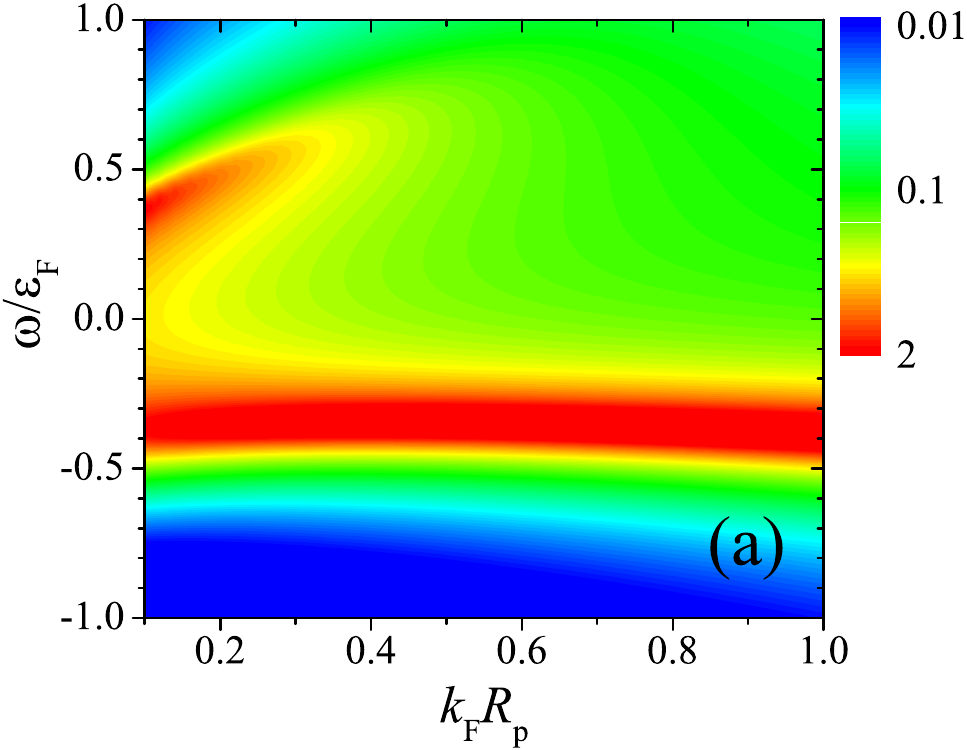}\includegraphics[width=0.475\textwidth]{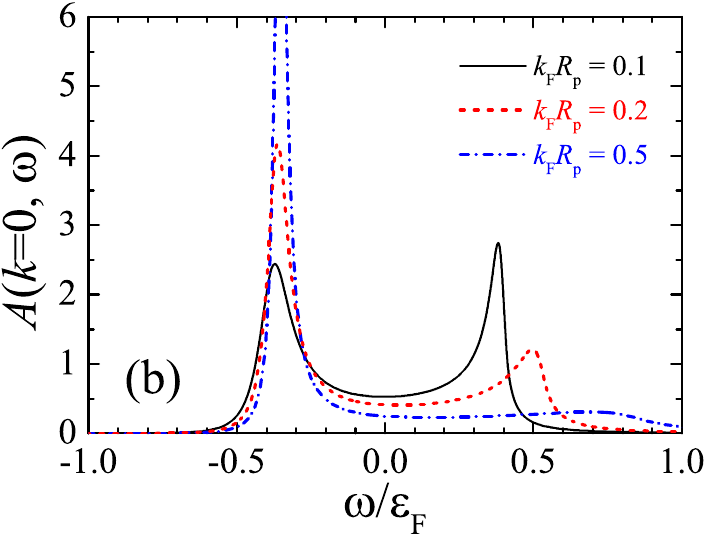}
\par\end{centering}
\centering{}\caption{\label{fig:fig5_AkwT020TF} Zero-momentum spectral function $A(k=0,\omega)$
of the impurity as a function of the interaction range $k_{F}R_{p}$
in the unitary limit $1/(k_{F}^{2}a_{p})=0$. The two-dimensional
contour plot in (a) is shown in the logarithmic scale, in units of
$\varepsilon_{F}^{-1}$. The three lines in (b) show the spectral
functions at $k_{F}R_{p}=0.1$ (black solid), $k_{F}R_{p}=0.2$ (red
dashed) and $k_{F}R_{p}=0.5$ (blue dot-dashed). Here, the temperature
is set to $T=0.2T_{F}$.}
\end{figure*}

To complete our analysis of the repulsive polaron in the unitary limit,
we show in Fig. \ref{fig:fig4_PolaronkFRp010} the temperature dependence
of the energy (in the main figure) and the decay rate (in the inset)
of both attractive Fermi polaron and repulsive Fermi polaron. As illustrated
by the red stars in the inset, the decay rate of the attractive Fermi
polaron rapidly increases with increasing temperature. It becomes
larger than the Fermi energy when $T>0.3T_{F}$, in agreement with
the observation in Fig. \ref{fig:fig1_akwkFRp010}(a) that the attractive
polaron ceases to exist at this temperature. On the other hand, as
indicated by black dots in the inset, the decay rate of the repulsive
Fermi polaron increases from $0.1\varepsilon_{F}$ at $T\sim0$ to
about $0.5\varepsilon_{F}$ at $T\sim0.7T_{F}$. By further increasing
temperature, the decay rate only slightly decreases, with a polaron
energy almost fixed at $0.27\varepsilon_{F}$ (see the main figure).
This provides a quantitative measure of the thermal robustness of
the repulsive Fermi polaron in the unitary limit.

\begin{figure}
\begin{centering}
\includegraphics[width=0.5\textwidth]{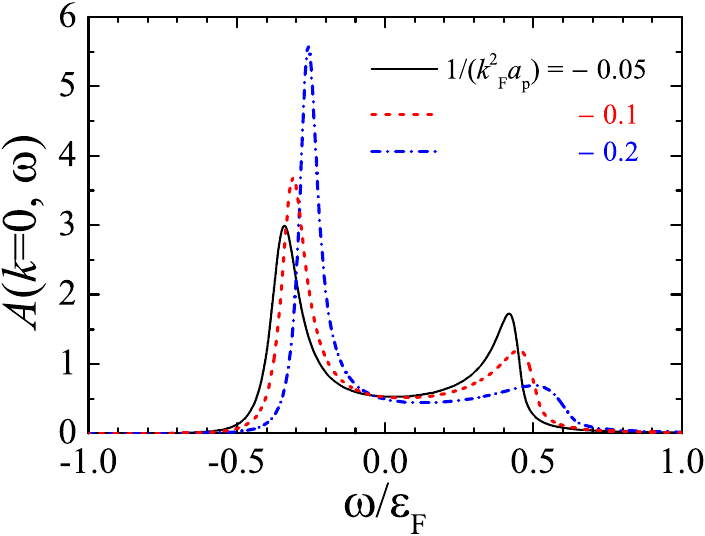}
\par\end{centering}
\centering{}\caption{\label{fig:fig6_AkwVaryInvkF2ap} Zero-momentum spectral function
$A(k=0,\omega)$ of the impurity at different inverse scattering areas:
$1/(k_{F}^{2}a_{p})=-0.05$ (black solid), $1/(k_{F}^{2}a_{p})=-0.1$
(red dashed) and $1/(k_{F}^{2}a_{p})=-0.2$ (blue dot-dashed). Here,
we take an interaction range $k_{F}R_{p}=0.1$ and set the temperature
$T=0.2T_{F}$.}
\end{figure}

\subsection{Parameter space for repulsive Fermi polarons}

We now turn to explore the parameter window of repulsive Fermi polarons,
for a $p$-wave interaction strength that does not support a two-body
bound state. We focus on the cases of a fixed temperature $T=0.2T_{F}$,
but with varying effective interaction range $k_{F}R_{p}$ and with
varying scattering area $1/(k_{F}^{2}a_{p})$.

In Fig. \ref{fig:fig5_AkwT020TF}, we report the evolution of the
zero-momentum spectral function $A(\mathbf{k}=0,\omega)$ as the interaction
range $k_{F}R_{p}$ increases in the unitary limit. From the two-dimensional
contour plot in Fig. \ref{fig:fig5_AkwT020TF}(a), we find that the
lower attractive polaron branch becomes increasingly dominant with
respect to the increase in the interaction range. However, the interesting
new feature of repulsive polaron branch quickly disappears when $k_{F}R_{p}$
becomes larger than $0.3$. At the interaction range $k_{F}R_{p}=0.5$,
as shown in Fig. \ref{fig:fig5_AkwT020TF}(b) by the blue dot-dashed
line, one can observe an extremely broad bump at about $\omega\sim0.7\varepsilon_{F}$,
as a reminiscent of the repulsive polaron. Therefore, we conclude
that a large interaction range does not favor the formation of a repulsive
Fermi polaron.

In Fig. \ref{fig:fig6_AkwVaryInvkF2ap}, we present the zero-momentum
spectral function at three inverse scattering areas $1/(k_{F}^{2}a_{p})=-0.05$
(black solid line), $-0.1$ (red dashed line) and $-0.2$ (blue dot-dashed
line), and at a fixed interaction range $k_{F}R_{p}=0.1$. By decreasing
inverse scattering area, the attractive Fermi polarons show a blue-shift
in energy. More importantly, the attractive polaron peak becomes sharper.
The repulsive Fermi polarons also show a blue-shift in energy. However,
their width becomes much wider with increasing inverse scattering
area. At $1/(k_{F}^{2}a_{p})=-0.2$, we may hardly identify the peak
as a well-defined repulsive polaron. This finding is consistent with
the earlier observation from Fig. \ref{fig:fig3_moleculeAkw}(a) that
the quasi-bound dressed molecule state becomes less well-defined as
the inverse scattering area decreases and moves away from the Feshbach
resonance.

\section{Conclusions and outlooks}

In conclusions, we have investigated $p$-wave Fermi polarons in two
dimensions at finite temperature, which potentially can be experimentally
realized in a population imbalanced spin-1/2 Fermi gas, where minority
atoms in one hyperfine state act as impurities and interact with majority
atoms in another hyperfine state near a $p$-wave Feshbach resonance.
In contrast to the conventional $s$-wave case that the existence
of a repulsive Fermi polaron requires a two-body bound state \cite{Scazza2017,Massignan2011,Hu2022},
a $p$-wave repulsive Fermi polaron can arise in the absence of two-body
bound state near the Feshbach resonance, due to a quasi-bound dressed
(many-body) molecule state that is hidden inside the two-particle
scattering continuum. The $p$-wave repulsive polaron shows a remarkable
stability against temperature. This extraordinary thermal robustness
would be very useful for its experimental observation, since a $p$-wave
Fermi gas is not blessed by Pauli exclusion principle and often has
severe atom loss below the Fermi temperature for degeneracy \cite{Luciuk2016,Yoshida2018}.

Instead of using a spin-1/2 Fermi gas with two hyperfine states, one
may also consider a mass-imbalanced Fermi-Fermi mixture such as $^{6}$Li-$^{40}$K
atomic mixture near Feshbach resonances, with a strong atom-dimer
attraction occurring between $^{40}$K atoms and weakly-bound $^{6}$Li-$^{40}$K
molecules in odd partial-wave channels \cite{Jag2014}. This higher
partial-wave attraction is mainly $p$-wave and is recombination-free
(and therefore stable), as experimentally observed \cite{Jag2014}.
We may tune down the number of weakly-bound $^{6}$Li-$^{40}$K molecules
to treat them as impurities. In this case, the mass of impurity is
slightly larger than the mass of the fermions in the Fermi sea. Our
results of repulsive Fermi polarons, based on equal mass, should be
qualitatively applicable. Quantitative predictions however require
the extension of our work to account for an arbitrary impurity mass.
Another crucial issue of the atom-dimer $p$-wave attraction is that
weakly bound dimers may spontaneously dissociate on a time scale of
about tens mill-seconds \cite{Jag2014}. We would like to leave a
careful investigation of these two issues (i.e., the unequal mass
effect and the short lifetime of impurity) to a future study. 

\section{Statements and Declarations}

\subsubsection{Ethics approval and consent to participate }

Not Applicable.

\subsubsection{Consent for publication }

Not Applicable.

\subsubsection{Availability of data and materials }

The data generated during the current study are available from the
contributing author upon reasonable request. 

\subsubsection{Competing interests}

The authors have no competing interests to declare that are relevant
to the content of this article. 

\subsubsection{Funding}

This research was supported by the Australian Research Council's (ARC)
Discovery Program, Grants No. FT230100229 (J.W.).

\subsubsection{Authors' contributions }

All the authors equally contributed to all aspects of the manuscript.
All the authors read and approved the final manuscript. 

\subsubsection{Acknowledgements }

The present work is dedicated to the memory of Professor Lee Chang,
whose contributions to physical science and education were longstanding
and far-reaching. He enthusiastically carried out the research on
ultracold atomic physics in Tsinghua University twenty-five years
ago and guided the authors (HH and XJL) into this fantastic field.

\subsubsection{Authors' information}

Hui Hu, Centre for Quantum Technology Theory, Swinburne University
of Technology, Melbourne 3122, Australia, Email: hhu@swin.edu.au

Jia Wang, Centre for Quantum Technology Theory, Swinburne University
of Technology, Melbourne 3122, Australia, Email: jiawang@swin.edu.au

Xia-Ji Liu, Centre for Quantum Technology Theory, Swinburne University
of Technology, Melbourne 3122, Australia, Email: xiajiliu@swin.edu.au

\appendix
\begin{figure}
\begin{centering}
\includegraphics[width=0.5\textwidth]{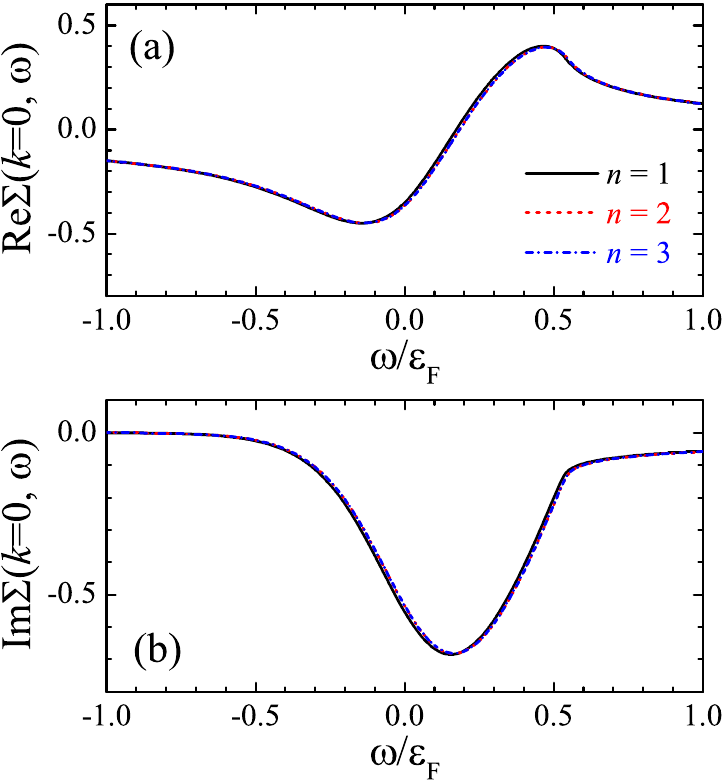}
\par\end{centering}
\caption{\label{fig:fig7_ndep} Impurity self-energy (in units of $\varepsilon_{F}$)
calculated at different values of $n$, as indicated. Here, we consider
the unitary limit $1/(k_{F}^{2}a_{p})=0$ and take an interaction
range $k_{F}R_{p}=0.20$. The temperature is $T=0.2T_{F}$.}
\end{figure}

\section{The $n$-dependence of the separable interaction form}

Throughout the work, we have taken $n=1$ in our separable interaction
form $V_{kk'}=$$\lambda\Gamma(k)\Gamma(k')$, with a $p$-wave form
factor $\Gamma(k)=(k/k_{F})[1+(k/k_{0})^{2n}]^{3/2}$. After the renormalization,
physical results (at low-energy) should only depend on the scattering
area $1/(k_{F}^{2}a_{p})$ and interaction range $k_{F}R_{p}$, \emph{independent}
on the choice of the value of $n$. In Fig. \ref{fig:fig7_ndep},
as an example we explicitly examine this independence for the impurity
self-energy in the unitary limit $1/(k_{F}^{2}a_{p})=0$. We find
negligible differences when we change the value of $n$ from 1 to
3, as anticipated.

\section{The numerical calculation of $T_{2}^{-1}(\mathbf{q},\omega)$}

In numerical calculations, it is convenient to take the natural units,
where $m=\hbar=k_{B}=1$. In other words, we set the units of energy
and momentum as $\varepsilon_{F}$ and $k_{F}$, respectively. The
inverse of the many-body $T$-matrix then takes the form,
\begin{equation}
T_{2}^{-1}\left(q,\omega\right)=\frac{1}{\lambda}+\intop_{0}^{\infty}\frac{pdp}{4\pi}\frac{g\left(p\right)}{p^{2}-\left(\frac{\omega+\mu}{2}-\frac{q^{2}}{4}\right)-i0^{-}},\label{eq:invT2B1}
\end{equation}
where we have defined an angle-integrated function,
\begin{equation}
g\left(p\right)\equiv\Gamma^{2}\left(p\right)\intop_{0}^{2\pi}\frac{d\varphi}{2\pi}f\left[-\left(p^{2}+\frac{1}{4}q^{2}+pq\cos\varphi-\mu\right)\right].
\end{equation}
The integral Eq. (\ref{eq:invT2B1}) is well defined if $y\equiv(\omega+\mu)/2-q^{2}/4<0.$
In this case, we find that,
\begin{eqnarray}
\textrm{Re}T_{2}^{-1} & = & \frac{1}{\lambda}+\intop_{0}^{\infty}\frac{dp}{4\pi}\frac{pg\left(p\right)}{p^{2}+\left|y\right|},\\
\textrm{Im}T_{2}^{-1} & = & 0.
\end{eqnarray}
Otherwise ($y\geq0$), we may use the identity
\begin{equation}
\frac{1}{X-i0^{+}}=\frac{\mathscr{P}}{X}+i\pi\delta\left(X\right)
\end{equation}
to recast the real and imaginary parts of $T_{2}^{-1}(q,\omega)$
into the forms,
\begin{eqnarray}
\textrm{Re}T_{2}^{-1} & = & \frac{1}{\lambda}+\frac{\mathscr{P}}{4\pi}\intop_{0}^{\infty}dp\frac{pg\left(p\right)}{p^{2}-y}=\frac{1}{\lambda}+\frac{C_{1}+C_{2}}{8\pi},\\
\textrm{Im}T_{2}^{-1} & = & \frac{1}{4\pi}\pi\intop_{0}^{\infty}dppg\left(p\right)\delta\left(p^{2}-y\right)=\frac{g\left(\sqrt{y}\right)}{8}.
\end{eqnarray}
 Here, by taking Cauchy principle value we have defined two integrals,
\begin{eqnarray}
C_{1} & \equiv & \intop_{0}^{\infty}d\xi\frac{g\left(\sqrt{2y+\xi}\right)}{y+\xi},\\
C_{2} & \equiv & \intop_{0}^{y}d\xi\frac{g\left(\sqrt{y+\xi}\right)-g\left(\sqrt{y-\xi}\right)}{\xi}.
\end{eqnarray}

\end{document}